\documentclass[pre,twocolumn,floatfix,superscriptaddress,a4paper, amsmath,amssymb]{revtex4-2}

\usepackage{hyperref}
\usepackage[section]{placeins}

\usepackage{graphicx}
\usepackage{bm}

\usepackage{subfiles} 

\usepackage{xcolor}

\usepackage[utf8]{inputenc}

\usepackage{subfiles} 
\usepackage{soul}
\usepackage{gensymb}

\usepackage{tikz}

\begin{document}

\title{Confinement-induced motion of ciliates}

\author{G. C. Antunes}
 \email{g.antunes@tu-berlin.de}
\affiliation{Institute of Physics and Astronomy,
Division of Theoretical Physics, Technische Universit\"at Berlin, Hardenbergstrasse 36,
10623, Berlin, Germany}

\author{C. Obst}
\affiliation{Institute of Physics and Astronomy,
Division of Theoretical Physics, Technische Universit\"at Berlin, Hardenbergstrasse 36,
10623, Berlin, Germany}

\author{H. Stark}
\affiliation{Institute of Physics and Astronomy,
Division of Theoretical Physics, Technische Universit\"at Berlin, Hardenbergstrasse 36,
10623, Berlin, Germany}

\date{\today}

\begin{abstract}
The time dynamics of flagellar and ciliary beating is often neglected in theories of microswimmers, with the most common models prescribing a time-constant actuation of the surrounding fluid. By explicitly introducing a metachronal wave, coarse-grained to a sinusoidal surface slip velocity, we show that a 
spatial resonance between the metachronal wave and the corrugation of a confining cylindrical channel enables a ciliate to swim even when it cannot move forward in a bulk fluid. 
Using lubrication theory, we reduce the problem to the Adler equation that reveals an oscillatory and steady swimming regime. Interestingly, a ciliate can even reverse its swimming direction in a corrugated channel compared to the bulk fluid. 
\end{abstract}

\maketitle


\begin{figure}[t]
	\centering
   \includegraphics[width=1\columnwidth]{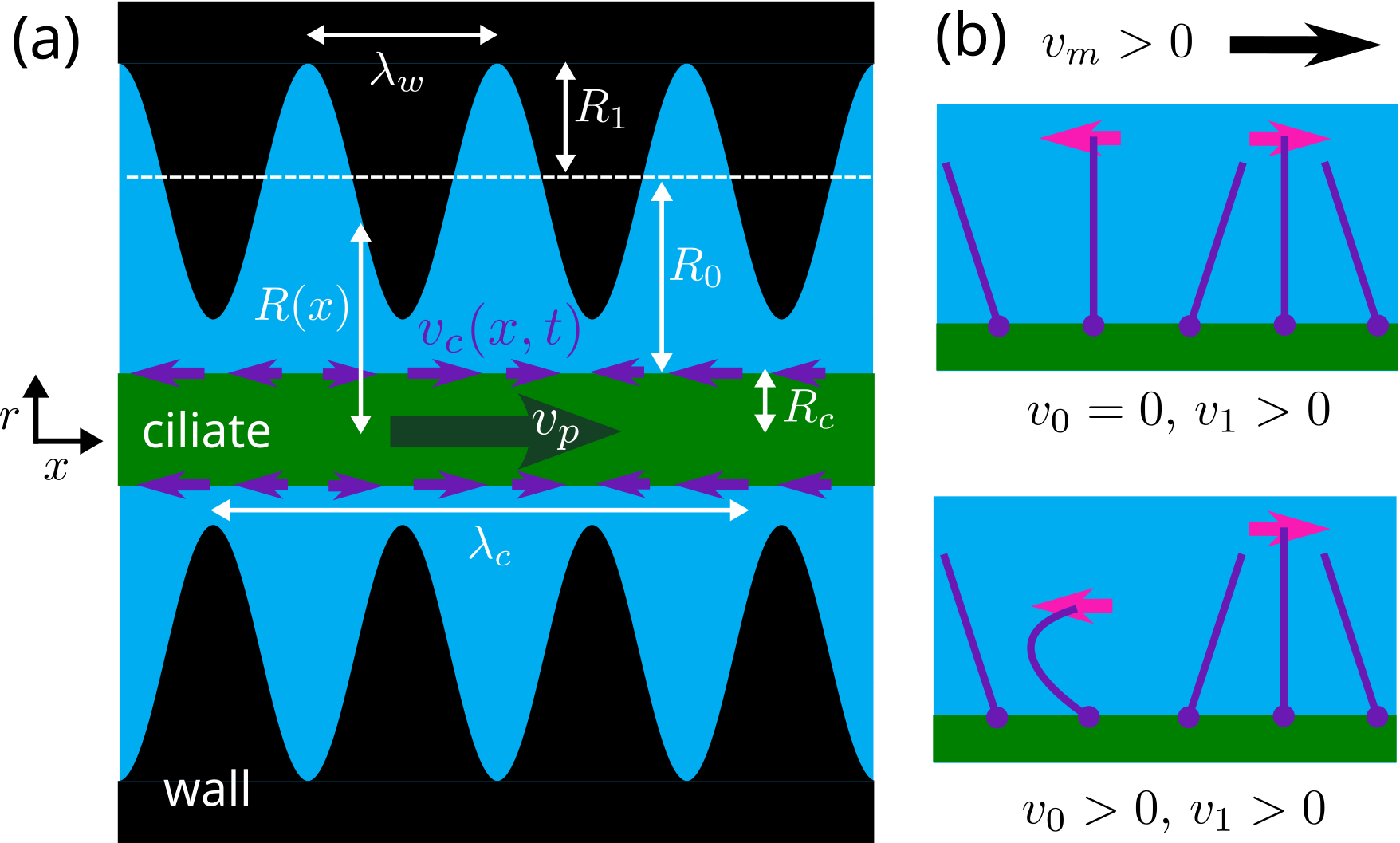}
   \caption{(a) 
   Schematic of a long ciliate (green cylinder) that swims in a cylindrical channel with corrugated walls, which are depicted in black while the surrounding fluid is drawn in blue. The cilia are coarse-grained to an effective slip velocity $v_c$, represented by purple 
   arrows on the surface of the ciliate, while the ciliate moves with velocity $v_p$. (b) Schematics of a reciprocal (above) and non-reciprocal (below) ciliary beat pattern. Pink arrows indicate the direction of motion of the single cilia. The metachronal wave travels from left to right with speed $v_m >0$. For a sufficiently sparse ciliary carpet, the net fluid flow of the reciprocal beat pattern is negligible. 
   } 
        	\label{fig:model}
\end{figure}

It is widely known that bodies undergoing a reciprocal motion cannot propel in bulk fluids at low Reynolds numbers \cite{Purcell1977}. Thus, microorganisms need to rely on structures such as flagella \cite{Mukherjee2014} or cilia \cite{Lin2014} which beat in a non-reciprocal fashion \cite{Purcell1977,Lauga2011}.
Due to the challenge of fully capturing the dynamics of each flagellum or cilium, theoretical descriptions of microswimmers typically rely on a coarse-graining. Examples include the specification
of an effective surface slip velocity in the well-known squirmer model \cite{Lighthill1952,Blake1971,Ishikawa2024, Chisholm2016, Zantop2020,Ganguly2025} 
or a truncated multipolar expansion for the swimmer-generated flow fields
\cite{Lauga2012,Lauga2016,Ishikawa2024, Yeomans2014,Drescher2010}. Such models have been successful in shedding light on the behavior of microswimmers in complex environments, including swimmer
suspensions \cite{Pagonabarraga2013, Alarcon2013, Shum2025, Ishikawa2007, Ishikawa2008,Yoshinaga2017}, 
also under confinement by flat \cite{Ahana_2019,WuZhang2024, HernandezOrtiz_2009, Llopis2010,Ishimoto2013,Ruhle2018, Shen2018,Zantop2020,Ruhle2020,Lintuvuori2016} or curved \cite{Kurzthaler2021,Ishimoto2023,Kuron2019,Chaithanya2021, Ishikawa2018, Barri2022,Zheng2023} walls, 
as well as
porous \cite{Chamolly2017,Kamarapu2022} and deformable \cite{Shaik2017} materials.

However, nearly all of the cited theoretical work focuses on the time-averaged actuation of the fluid by the flagellar or ciliary beat, \emph{i.e.}, prescribes slip velocity profiles and multipolar expansion coefficients that are constant in time.
While such descriptions may capture the dynamics of microswimming on long timescales, they fail to describe the dynamics on the timescale of the beating period \cite{Klindt2015,Pedley2016,Guasto2010,Redaelli2023}. In this Letter, we show that accounting for the time dynamics of ciliary beating is crucial to understanding the motion of ciliates in spatially inhomogeneous confinement. In particular, their ability to move forward 
is heavily effected by a spatio-temporal resonance between the fluid forcing 
coming from the ciliary beating pattern and the modulated environment.

In our studies, we use both lattice-Boltzmann simulations and a theoretical analysis based on the lubrication approximation. The latter is a versatile tool and has been used extensively in different contexts such as biological flows \cite{Lubrication_Stegemerten2022,Ureter_Zheng2021,Ureter_Lykoudis1970,Kozlov2015}, fluid-mediated interactions \cite{Dowson1962, Patel2022,Davis1986,Rallabandi2024},  chemically-active \cite{Antunes2022, Antunes2023, Antunes2024,Lubrication_Richter2025,Lubrication_Voss2024} or charged \cite{Bai2006,Malgaretti2014} flows, as well as droplets and films \cite{Lubrication_Thiele2003, Lubrication_Witelski2000,Lubrication_Yang2025,Lubrication_Charitatos2020}.

In the following, we demonstrate how a sinusoidal metachronal wave coarse-grained to a surface slip velocity that spatially resonates with the corrugation of a confining channel enables ciliates to swim, even when they cannot move forward in a bulk fluid. In particular, we reduce the swimming dynamics to the Adler equation that shows an oscillatory and a steady regime.

Specifically, we aim to investigate the motion of a ciliate in a cylindrical channel aligned with the $x$ axis \footnote{In our work, we consider straight channels to bring out the basic mechanism of the novel phenomenology. However, it also applies
to channels with curvature radii
much larger than the dimensions of the ciliate.}
 and whose local radius $R(x)$ varies according to
\begin{equation}
    \label{eq:Radius}
    R(x) = R_c + R_0 + R_1 \cos(k_w x) \, .
\end{equation}
Here, $R_c$ is the radius of the ciliate, $R_0$ is the spatially averaged distance between the ciliate and the channel wall, 
$R_1$ is the amplitude of the corrugation, and $k_w$ is its wavenumber [see Fig. \ref{fig:model} (a)]. The ciliate is modeled as a cylinder whose length $L_c$ is much larger than the wavelength $\lambda_w = 2\pi/k_w$ and so can be considered effectively infinite in length. 
Ciliates are covered by a carpet of beating cilia, which move fluid along the ciliate surface and thereby propel the microorganism forward. Because variations in height of the ciliary layer are typically much smaller than $R_c$ \cite{Paramecium_Jana2012}, we coarse-grain
the operation of this layer into an effective slip velocity tangential to the surface of the ciliate,
\begin{equation}
    \mathbf{v}(x,r=R_c,t) = [v_p(t) + v_c(x,t)] \mathbf{e}_x,
    \label{eq_active_bc}
\end{equation}
Here, $\mathbf{v}(x,r,t)$ is the velocity of the fluid constrained between the ciliate and the corrugated channel wall, $v_p(t)$ is the swimming velocity of the ciliate, which needs to be determined, and the slip velocity
$v_c(x,t)$ encodes the ciliary beating. Contrary to previous work, we do not take the velocity $v_c(x,t)$ to be constant in time, as we aim to capture the time-varying actuation by the ciliary layer and its effect on the surrounding fluid. Indeed, cilia beat in a periodic fashion with period $T_c$ and their motions
often split into a power and recovery stroke \cite{Gray1922, Lin2014} [see Fig.\ \ref{fig:model}(b)] that are non-reciprocal, enabling net swimming and motion of fluid.

Typically, neighboring cilia beat with a phase lag and thereby form a metachronal wave \cite{Brumley2012,Guirao2007,Poon2023,Elgeti2013} 
with a wavelength $\lambda_c$ much smaller than the size $L_c$ of the ciliate \cite{Paramecium_Jana2012}. To capture these features, we propose for the slip velocity at the surface of the ciliate,
\begin{equation}
    \label{eq:slip_velocity}
    v_c(x,t) = v_0 + v_1 \cos\{k_c [x-x_p(t)] - \omega t\},
\end{equation}
where $v_0$ is the time-averaged velocity imposed by the ciliary layer, and $v_1$ is the amplitude of the metachronal wave. Here, $v_0 \neq 0$ indicates that the ciliary carpet induces a net fluid transport across the surface of the ciliate. Such a transport may result from non-reciprocal power and recovery strokes or even from reciprocally beating cilia that
are packed densely enough so that hydrodynamic interactions lead to net fluid flow \cite{Khaderi2012}.
Furthermore, $k_c = 2\pi/\lambda_c$ is the wavenumber of the metachronal wave and $\omega = 2\pi/T_c$ is the ciliary frequency. As a result, the metachronal wave travels with a speed $v_m=\omega/k_c$ in the body-fixed frame of the ciliate. The 
origin of this frame of reference moves with velocity $v_p(t)$ along the $x$ direction and is located at position $ x_p(t) = \int_0^t v_p(t^{\prime}) dt^{\prime}$.

Now, the main task is to determine the swimming velocity of the ciliate, $v_p(t)$. At low Reynolds numbers, the fluid obeys Stokes' equations and the incompressibility condition,
 $\nabla P(\mathbf{r},t) = \mu \Delta \mathbf{v}(\mathbf{r},t)$ and $ \nabla \cdot \mathbf{v}(\mathbf{r},t) = 0 $, where $\mu$ is the dynamic viscosity and $P(\mathbf{r},t)$ is the pressure. No-slip boundary conditions apply at the surface of the channel, $  \mathbf{v}(x,r=R(x),t) = 0 $. The ciliate swims in an otherwise quiescent fluid such that no external pressure gradient is applied, and moves force-free meaning without an applied external force, $ \oint_S \boldsymbol{\sigma}(\mathbf{r},t) \cdot \mathbf{n}(\mathbf{r},t)  dS = \mathbf{0} 
$, 
where $S$ is the surface of the ciliate, $\boldsymbol{\sigma}(\mathbf{r},t)$ is the Newtonian stress tensor, $    \boldsymbol{\sigma}(\mathbf{r},t) = -P(\mathbf{r},t) \mathbf{I} + \mu [\nabla \otimes \mathbf{v}(\mathbf{r},t) + (\nabla \otimes \mathbf{v})^T(\mathbf{r},t)]   $,
and $\mathbf{n}(\mathbf{r},t)$ is the outer normal. 

\begin{figure}
	\centering
   \includegraphics[width= 1 \columnwidth]{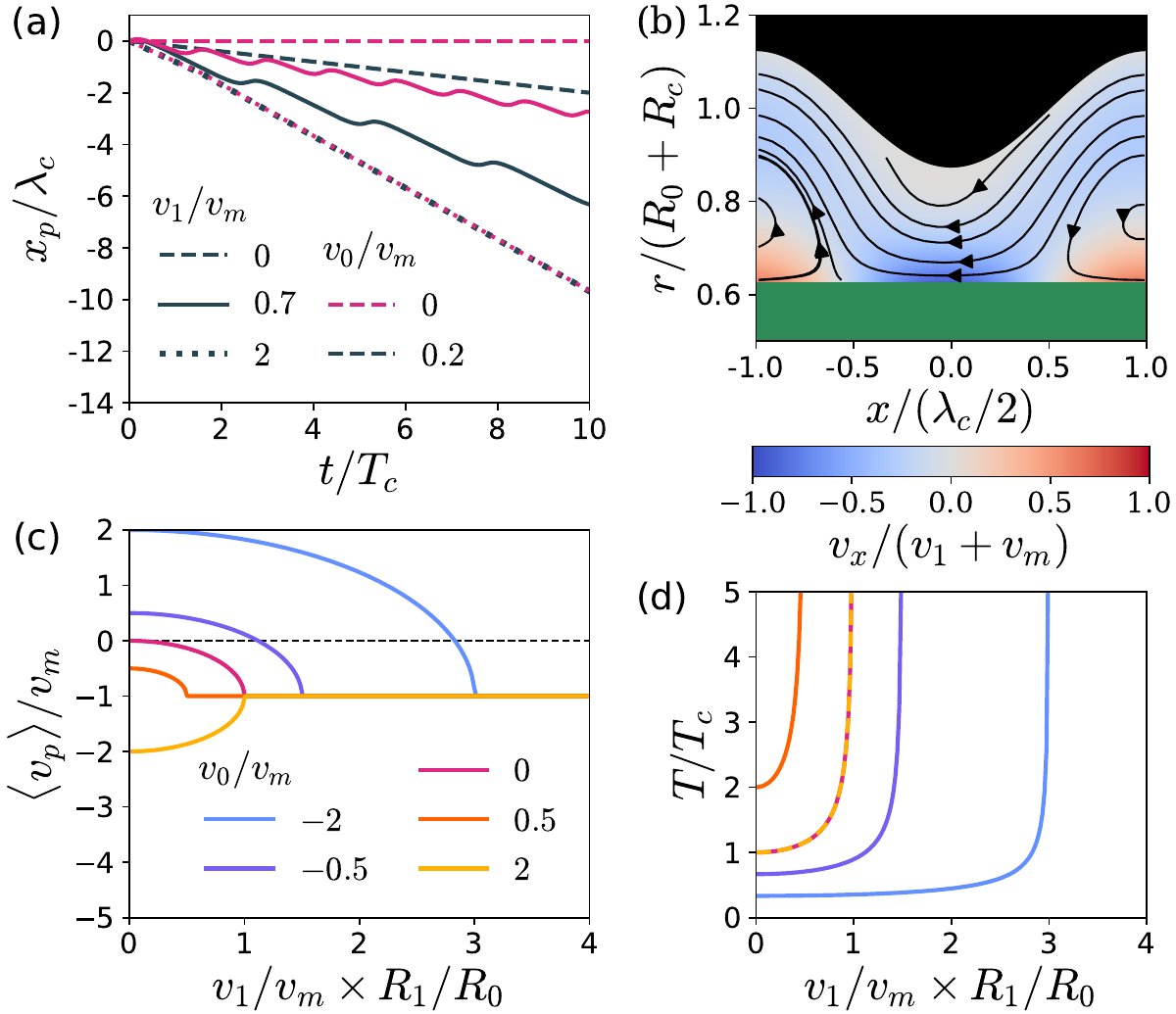}
   \caption{Resonance between confinement and metachronal wave. (a) Position $x_p$ of the ciliate plotted \emph{vs.} time for different values of $v_0/v_m$ and $v_1/v_m$ with $\mathcal{G}_n=2$. Colors indicate value of $v_0/v_m$ and line types indicate $v_1/v_m$. (b) Snapshot of the fluid flow for $v_0=0$, $R_0/R_c = 3/5$, $R_1/R_0=1/3$, $R_0/\lambda_c=3/20$, $v_1/v_m=3.6$, taken at $t/T_c = 3.78$.  
    The fluid streamlines are shown in black and the color indicates the flow-velocity component $v_x$. In this moment, the ciliate 
    moves to the left. (c) Time-averaged ciliate velocity $\langle v_p \rangle$ and (d) period of oscillatory motion $T$ plotted 
     \emph{vs.}
     the amplitude of ciliary slip velocity, $v_1/v_m$, times the amplitude of channel corrugation, $R_1/R_0$, for several values of $v_0/v_m$. In all panels $\lambda_c = \lambda_w$. 
}
    	\label{fig:purely_lubrication}
\end{figure}

\emph{Lubrication theory and linearization}.
We solve the above equations by employing the lubrication approximation \cite{Schlichting1979}, whereby the velocity field is taken 
to vary much more rapidly in the radial direction than in the $x$ direction. Such an assumption is true if the characteristic lengths 
in the radial direction ($R_0$, $R_1$) are much smaller than the characteristic lengths along the $x$ axis ($\lambda_w, \lambda_c$). 
Ultimately, we arrive at the following equation for the time evolution of $x_p(t)$, 
\begin{equation}
  \dot{x}_p(t) = -\langle \mathcal{G}(x) v_c(x,t) \rangle \, ,
 \label{eq_withMobility}
 \end{equation}
where the angle brackets indicate an average over $x$, and the local mobility coefficient $\mathcal{G}(x)$ only depends on the shape 
of the channel wall and the radius of the ciliate (see Supplemental Material in Sect. I ). 
In particular, $\langle \mathcal{G}(x) \rangle =1$. Since $v_c(x,t)$ is a sinusoid [Eq. \eqref{eq:slip_velocity}], one immediately obtains
\begin{equation}
     \dot{x}_p(t) = 
  \begin{cases}
\displaystyle
 -v_0 - v_1 \frac{ \mathcal{G}_n}{2}\cos\left[  k_c x_p(t) + \omega t \right]  &\text{if }  \lambda_c=\lambda_w/n \\
- v_0 & \text{if } \lambda_c \neq \lambda_w/n \, , \label{eq_lubTheory}
  \end{cases}
\end{equation}
where $n$ is a natural number and $\mathcal{G}_n$ are the respective Fourier coefficients of $\mathcal{G}(x)$. 
In the limits of small curvature of the ciliate body ($R_0/R_c \ll 1$) and shallow 
corrugation ($R_1/R_0 \ll 1$), a linearization with respect to these parameters shows that only the 
Fourier coefficient for $n=1$ is non-zero, becoming $\mathcal{G}_1 = 2R_1/R_0$. Clearly, in the case where the spatiotemporal dynamics of the ciliary 
carpet is discarded ($v_1=0$), the channel is flat ($R_1 =0$), or the wavelengths of the metachronal wave and the corrugation do 
not match ($\lambda_c \neq \lambda_w/n$), the ciliate swims with a velocity $v_p = - v_0$, as it would when placed in bulk fluid. 
However, if none of the three previous conditions are met, a resonance between the local hydrodynamic resistance (resulting from 
the ccorrugated wall profile) and the fluid forcing (resulting from the metachronal wave) generates the extra term in Eq.\ \eqref{eq_lubTheory}. 
The effect on the trajectory of the ciliate can be seen by numerically solving Eq.\ \eqref{eq_lubTheory}, the results of which we show 
in Fig.\ \ref{fig:purely_lubrication}(a). When $v_1/v_m =0.7$ (full lines), where $v_m$ is the metachronal-wave speed, the ciliate swims with an overall drift and superimposed oscillations.

Remarkably, such a drift occurs even when the ciliary carpet does not generate net fluid transport along its surface [$v_0=0$, magenta lines
 in Fig.\ \ref{fig:purely_lubrication}(a)]. An organism with $v_0=0$ cannot propel itself in a bulk fluid, but it is able to when placed in a 
 corrugated channel of appropriate wavelength. The cause is an imbalance between the net propulsion generated by the power stroke and 
 the recovery stroke. While both generate the same slip velocity (in magnitude), they are performed in regions of different film thickness, 
 where the generated flow experiences different effective hydrodynamic resistances [see Fig.\ \ref{fig:purely_lubrication}(b)]. The result is 
 directed motion even when the time-and-space-averaged slip velocity is zero. Every pair of forward and backward beating cilia can 
 be thought of as two people playing tug-of-war. 
Even if
 they are equally strong, the pair will still move if one person stands on dry concrete and the other on slippery mud. 
Thus, motion is
 enabled by a variation in the friction profile, which occurs on a length scale comparable to the rope's length. Depending on the phase 
 difference between the corrugated 
 wall and the velocity profile on the surface of the ciliate, the ciliate swims either with or against the metachronal wave. What enables net 
propulsion (rather than oscillation about a certain position) is the fact that the metachronal wave has a given direction. Because the 
velocity of the metachronal wave in the laboratory frame is the sum of the velocities of the ciliate velocity and of the metachronal wave, 
configurations of the flow leading to swimming against the wave persist longer than configurations which lead to swimming with the wave. 
As such, the ciliate spends more time swimming against the metachronal wave than with it, as can be seen in Supplemental Video 1 . 

For high enough values of 
$v_1$, the trajectory is no longer oscillatory, but instead becomes steady, in the direction opposite 
to the metachronal wave, and exactly at its speed [see Fig.\ \ref{fig:purely_lubrication} (a)].
Thus, no matter how strongly the ciliary carpet may push, the ciliate is constrained 
to swim with the metachronal phase speed $v_m$. In this regime, the slip velocity profile becomes stationary in the laboratory frame 
of reference, as can be seen in Supplemental Video 2 . Thus, the phase difference between the channel corrugation and the slip velocity profile is constant, and no further acceleration is possible.

One can indeed rationalize the two observed dynamical regimes by introducing the non-dimensional coordinate in the frame of reference 
moving with the metachronal wave,  $\xi = k_c x_p(t)  + \omega t + \pi/2$. Rescaling time by $\omega^{-1}$, Eq.\ (\ref{eq_lubTheory}) for $\lambda_c = \lambda_w/n$ can be cast in the form of the well-known Adler equation \cite{Strogatz2015, Adler1946, Grawitter2024},
\begin{equation}
\dot{\xi} = a - b\sin \xi \quad \text{with} \enspace a = 1- \frac{v_0}{v_m} \enspace \text{and} \enspace 
b =  \frac{v_1}{v_m}
\frac{\mathcal{G}_n}{2} \, ,
\label{eq_Adler}
\end{equation}
which indeed shows a dynamical transition for 
\begin{equation}
\left | \frac{a}{b} \right | = \left |  \left (1- \frac{v_0}{v_m}\right ) \Big/  \left ( \frac{v_1}{v_m} 
\frac{\mathcal{G}_n}{2}
\right) \right | = 1 \, .
\end{equation}
It is now possible to infer the time-averaged velocity of the ciliate, $\langle v_p \rangle$ \cite{Strogatz2015, Grawitter2024}. In the phase-locked regime ($|a/b| < 1$), one obtains $\langle v_p \rangle = - v_m $, 
while the oscillatory
regime ($|a/b| > 1$) yields
\begin{equation}
\frac{\langle v_p \rangle}{v_m} = \left(1 - \frac{v_0}{v_m} \right) 
\sqrt{1-\left(\frac{\frac{v_1}{v_m}\frac{\mathcal{G}_n}{2}}{1-\frac{v_0}{v_m}}\right)^2} -1  \, .
\label{eq.linearized}
\end{equation}

As mentioned earlier, in the linearized limit,
$\mathcal{G}_n = 2R_1/R_0 \, \delta_{1n}$, where $\delta_{ij}$ is the Kronecker symbol. 
So, in Fig.\ \ref{fig:purely_lubrication} (c), we plot the time-averaged ciliate velocity \emph{versus} $v_1/v_m \times R_1/R_0$ for 
different $v_0/v_m$ and $\lambda_c=\lambda_w$. For increasing 
$v_1R_1$,  the ciliate velocity approaches that of the metachronal wave. Depending on 
$v_0$, $\langle v_p \rangle$ either 
decreases or increases towards $-v_m$. In fact, the ciliate even reverses direction with respect to the bulk fluid if $v_0 /v_m < 0$ and 
$v_1/v_m \times R_1/R_0 > \sqrt{a^2-1}$.
Thus, an antiplectic ciliate, which swims in the same direction as the metachronal wave in the bulk fluid (\emph{e.g.} \textit{Paramecium} \cite{Funfak2014}), may become a symplectic swimmer, which moves opposite to the metachronal wave (\emph{e.g.} \textit{Opalina} \cite{Tamm1970}) in a sufficiently corrugated channel. 

The period of the oscillatory motion, $T$, can also be obtained analytically \cite{Strogatz2015, Grawitter2024},
\begin{equation}
T = \frac{T_c}{\sqrt{\left( \frac{v_0}{v_m} -1 \right)^2 - \left( \frac{v_1}{v_m}  
\frac{\mathcal{G}_n}{2}\right)^2}} \, .
\end{equation}
It diverges at the transition from the oscillatory to the phase-locked regime, and takes values both below and above $T_c$, as shown in Fig.\ \ref{fig:purely_lubrication}(d).

\begin{figure}[t]
	\centering
   \includegraphics[width= 1\columnwidth]{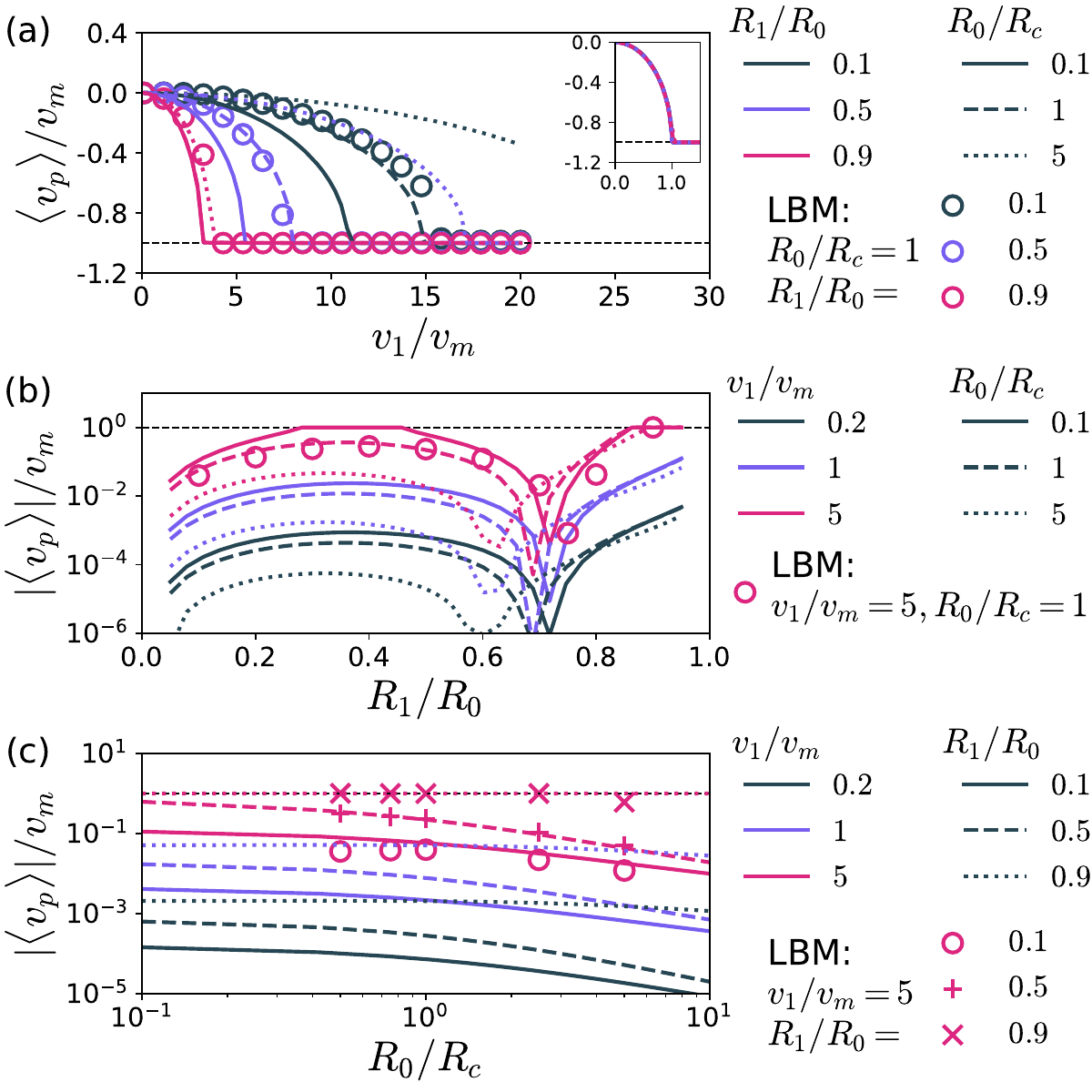}
   \caption{
   Confinement-induced motion. Time-averaged ciliate velocity $\langle v_p \rangle$ 
   as function of (a) ciliary wave amplitude $v_1/v_m$, (b) corrugation amplitude 
   $R_1/R_0$, and (c) ciliate curvature $R_0/R_c$ for $\lambda_c / \lambda_w=1$, $R_0/\lambda_c=1/4$, and $v_0 = 0$. The two varied parameters of the curves are indicated in the legends by color and line style, respectively. Colored symbols show results from lattice-Boltzmann simulations. 
   Inset in (a): master curve $\langle v_p \rangle / v_m = \sqrt{1-v_1^2/v_{*}^{2}} - 1$ with $v_* = 2v_m/\mathcal{G}_1$. 
   In (b), (c) the magnitude $|\langle v_p \rangle|$ is plotted. Colors indicate values of $R_1/R_0$ 
   in (a) and $v_1/v_m$ in (b), (c),
   while line types indicate values of 
   $R_0/R_c$ in (a), (b) and $R_1/R_0$ in (c).
         }
    	\label{fig:effect_R0Rc_R1R0}
\end{figure}


\emph{Full lubrication theory and lattice-Boltzmann simulations.}
In the full theory, the Fourier coefficients $\mathcal{G}_n$, and thereby $\langle v_p \rangle$, have a non-trivial dependence on 
$R_1$ and $R_c$, which we now explore.  We also support our findings with lattice-Boltzmann simulations \cite{Krueger_book}. 
In the following, we shall focus on the motion of ciliates with $v_0 =0$ since this case provides the clearest and physically most 
interesting manifestation of the interplay between corrugation and slip velocity.

First, in  Fig.\ \ref{fig:effect_R0Rc_R1R0}(a) we again observe the expected transition from the oscillatory to the phase-locked regime
with $\langle v_p \rangle = - v_m$ for increasing wave amplitude
$v_1$. As deduced from Eq.\ (\ref{eq.linearized}), all curves collapse on a master curve of the form  $\langle v_p \rangle / v_m = \sqrt{1-v_1^2/v_*^{2}}- 1$ with the critical wave amplitude 
$v_* = 2v_m/\mathcal{G}_1$ separating the two dynamical regimes from each other [see inset of Fig.\ \ref{fig:effect_R0Rc_R1R0}(a)].
In Fig.\ \ref{fig:effect_R0Rc_R1R0}(b), the swimming velocity shows two local maxima for increasing corrugation 
$R_1/R_0$. The maximum 
at $R_1/R_0 \approx 1$ corresponds to the limiting case where fluid transport through the bottlenecks and along the entire ciliate is greatly
suppressed such that the velocity field instead resembles an array of vortices. Swimming in such a channel geometry is not due to fluid 
transport from fore to aft, but due to an effective grip between the ciliate and the channel wall mediated by the enclosed pockets of fluid. 
Surprisingly, between the two local maxima, a strong dip in $|\langle v_p \rangle|$ occurs at 
$R_1^*/R_0$, the 
value of which is mainly determined by $R_0/R_c$.
In fact, the instantaneous ciliate velocity becomes zero as well due to
a root of the function $\mathcal{G}_1(R_1/R_0)$ 
(see the Supplemental Material, Sect. II ).
No matter the phase shift between the wall corrugation and the metachronal wave, the cilia beatings in the fore and aft directions 
compensate each other.
The dip in $v_p(t)$ arises when the system transitions between two flow morphologies. For $R_1/R_0$ 
below $R_1^*/R_0$,
the velocity field reveals a fluid layer that spans the 
entire modulated channel and flows in the same direction as the ciliate [blue-shaded region in Fig.\ \ref{fig:purely_lubrication} (b)]. 
In the opposite case (e.g. $R_1/R_0=0.8$), the 
size of this layer is greatly reduced, 
and
the flow is dominated instead by large vortices within the pockets. At $R_1/R_0=R_1^*/R_0$, 
the two flow types cancel each other and $\langle v_p \rangle$ drops to zero. More details can be found in the Supplemental Material together with illustrative Videos 4-6 . Finally, Fig.\ \ref{fig:effect_R0Rc_R1R0}(c) demonstrates that with increasing ciliate curvature 
$R_0/R_c$, the swimming velocity decreases. Thus, ciliates with negligible curvature ($R_0/R_c \rightarrow 0$) swim fastest. 
The presented curves are directly supported by lattice-Boltzmann simulations, as the symbols in Figs.~\ref{fig:effect_R0Rc_R1R0}(a)-(c) show.

\begin{figure}[t]
	\centering
   \includegraphics[width=1.0\columnwidth]{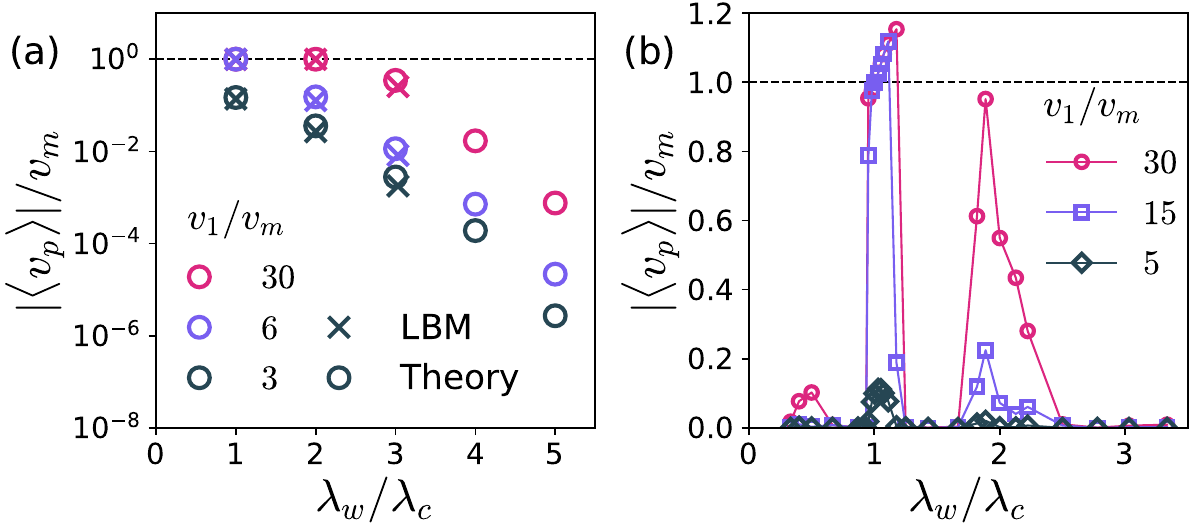}
   \caption{ 
   Resonance for finite ciliates. Time-averaged ciliate velocity $|\langle v_p \rangle|$ \emph{vs.} the ratio $\lambda_w/\lambda_c$ for various $v_1/v_m$. (a) Finite ciliate with length $L_c/\lambda_w \gg 1$. (b) Results from lattice-Boltzmann simulations for a finite, spherocylindrical ciliate with length $L_c / \lambda_c = 3.2$ and $d/L_c = 0.25$, where $d$ 
     is the surface-to-surface distance between the ciliate and its periodic image. For both panels, $R_0/R_c = 3/5$ and $R_1/R_0=1/3$. For panel (a) $R_0/\lambda_c=1/10$, and (b) $R_0/\lambda_w = 3/20$. }
    	\label{fig:resonance}
\end{figure}

Now, we relax the condition $\lambda_c = \lambda_w$ and show in Fig.\ \ref{fig:resonance}(a) that net swimming is also possible for the higher harmonics, $\lambda_w/\lambda_c =n$, as predicted by Eq.\ (\ref{eq_lubTheory}).
The swimming velocity decreases with this ratio once the phase-locked regime with $v_p/v_m = -1$ is no longer reached. 

Finally, we relax the limit of a very long ciliate, which allows swimming motion only at the metachronal wavelengths $\lambda_c=\lambda_w/n$. 
To demonstrate the corrugation-enabled motion also for finite ciliates, we perform lattice-Boltzmann simulations for ciliates shaped like spherocylinders with no-slip boundary conditions on the spherical half-caps, with a total length $L_c / \lambda_c = 3.2$ (Supplemental Video 3 ), 
as well as for ellipsoidal-shaped ciliates (Supplemental Videos 7 and 8 ). We again observe oscillatory trajectories with net drift
for relevant lengths $R_0/\lambda_c = R_0/\lambda_w = 1$, where lubrication theory is no longer valid.

As expected, peaks in the time-averaged propulsion velocity exist around the integer ratios $\lambda_w/\lambda_c$ but now they have a finite width
[see Fig.\ \ref{fig:resonance}(b)]. Further interesting features are visible: motion is also possible for $\lambda_w/\lambda_c < 1$, a time-averaged speed larger than $v_m$ occurs, and the peak for $\lambda_w/\lambda_c = 3$ is suppressed. However, a more thorough investigation of the influence of $L_c / \lambda_c$ on the propulsion speed is beyond the scope of this Letter. As a final test, our results remain valid when expanding our model to include a metachronal wave, which propagates oblique to the $x$ axis. In such a case, the dynamical regimes we describe in this article appear when the channel shows indentations in the azimuthal direction which corkscrew along the $x$ axis, similar to the internal surface of a rifle's barrel (Supplemental Material in Sec. III/ Supplemental Vi\-deo 9 ).

Using lubrication theory and lattice-Boltzmann simulations, we show that the explicit time dynamics of metachronal waves enables ciliates to swim in a corrugated channel, even if they do not move, on average, in a bulk fluid. This is due to a resonance between the local hydrodynamic resistance, arising from a spatially-varying radius of the channel, and the active flows generated by a sinusoidal metachronal wave. We 
rationalize our findings 
via the Adler equation that describes the transition from the oscillatory to 
the steady swimming regime. The mean swimming velocity can even reverse compared to the bulk fluid. 

This reversal of direction of motion may be used to fabricate a filter or chromatograph, which blocks the passage of a microorganism with a metachronal wave matching the channel corrugation, 
but allows other microorganisms to pass. Estimates for \textit{Paramecium} and a mutant show that the slow down and even reversal of the swimming velocity in a corrugated channel 
may be observable (see End Matter).
Despite motivating our model via ciliated microorganisms, our work provides a deeper understanding of how swimmers with a spatio-temporal fluid forcing at their surface behave in a corrugated channel, regardless of the nature of this forcing. The mechanism described in this Letter is in fact general enough that we expect it to also apply to other living beings, which propel via a metachronal wave of forcing, such as the gait displayed by centipedes \cite{Rieu2024}, milipedes \cite{Garcia_2021}, or cockroaches \cite{Weihmann2017}. 
For these terrestrial creatures, a modulated ground friction profile can play a role analogous to the spatially nonuniform channel radius. We further emphasize the applicability of our results to artificial swimmers/walkers which employ e.g. magnetically-actuated ciliary carpets \cite{Gu2020}, for which parameters like $v_1/v_0$ can be tuned. The mechanism described in this Letter thus opens a new avenue for environment-sensitive motility and environment-mediated control of robots moving in confinement.


\section*{Acknowledgments}
We thank Andrej Vilfan and Paolo Malgaretti for fruitful discussions.

\section*{Data availability statement}
The data for the current publication is available on Zenodo \cite{Zenodo}.
\\
\\

\bibliography{refs}

\section*{End Matter\\ Experimental predictions}
Despite the scarcity of experimental data for most ciliates
(particularly with respect to the metachronal parameters $\lambda_c$, 
$\omega$, and the velocity amplitude $v_1$),
we are able to predict the corrugation needed 
to reverse the
direction of motion
of
\textit{Paramecium}. The parameters $R_c$, $L_c$, $v_0$, $\lambda_c$, and $\omega$ \cite{Paramecium_Jana2012} are approximately known, 
and the parameter $v_1/v_0$ may be estimated as $v_1/v_0 \approx 3$ upon comparison with the cilia covering other organisms such 
as \textit{Tetrahymena} \cite{Akisato2025} and \textit{Plenaria} \cite{Rompolas2010}. We predict that with an estimated swimming velocity of 
$\approx 784 \mathrm{\mu m/s}$\cite{Funfak2014} (in a microchannel without corrugation), a \textit{Paramecium} in a corrugated channel of 
average radius $\approx 57  \mathrm{\mu m}$ ($R_0/R_c =1$ \cite{Paramecium_Jana2012}) will invert direction when $R_1/R_0 \gtrapprox 0.96 $.
The effect of the corrugation is, however, already seen when $R_1/R_0 = 0.9$, for which the 
swimming
velocity is $\approx 65\%$
of the value in the bulk fluid.

A more amenable case is that of a \textit{Paramecium} mutant showing a deformed ciliary beat, which leads to a smaller 
swimming
speed of $\approx 196 \mathrm{\mu m}$ \cite{Funfak2014}. Assuming
that only
 the average actuation 
 of
the fluid and not 
the beat amplitude
is affected,
yields $v_1/v_0 \approx 12$, and an inversion of 
the swimming
direction when $R_1/R_0 \gtrapprox  0.875$. Furthermore, we predict a velocity $\approx 65\%$ 
of the bulk-fluid value
for $R_1/R_0 \approx 0.35$ and $R_1/R_0 \approx 0.825$.

All these numbers show that the slow down and even reversal of the swimming velocity in a corrugated channel 
should be observable in an experiment.

\clearpage

\onecolumngrid
\setcounter{equation}{0}
\setcounter{figure}{0}
\renewcommand\theequation{S\arabic{equation}}
\renewcommand\thefigure{S\arabic{figure}}

\setcounter{page}{1}

\section*{Supplemental material}

\section{Lubrication theory}
\label{SM_lub}
In this section, we solve the problem of a long ciliate swimming in a corrugated cylindrical channel in the lubrication approximation regime. In this regime, characteristic lengths ruling the dynamics along the channel in the $x$-direction
($\lambda_c, \lambda_w$) are much larger than the ones ruling the dynamics in the radial $r$-direction
($R_0, R_1$). We refer to Fig.\ \ref{fig:model}(a) in the main text for the definition of all the lengths.
We now define the dimensionless quantities 
\begin{equation}
    \hat{x} = \frac{x}{\overline{\lambda}}, \enspace \text{and} \enspace \hat{r} = \frac{r}{R_0}, 
\end{equation}
where $\overline{\lambda}$ is the smallest longitudinal characteristic length
\begin{equation}
    \overline{\lambda} = \text{min}(\lambda_c,\lambda_w).
\end{equation}
Note that because $R(x)$ is greater than zero, we have $R_0 >R_1$, and so $R_0$ is the largest transversal characteristic length. In cylindrical coordinates the $x$-component of the Laplacian of the velocity vector in the Stokes equations can then be written as
\begin{equation}
    (\Delta \mathbf{v} )\cdot \mathbf{e}_x = \frac{1}{R_0^2}\frac{1}{\hat{r}} \partial_{\hat{r}} (\hat{r} \partial_{\hat{r}} v_x) + \frac{1}{\overline{\lambda}^2}\partial_{\hat{x}}^2 v_x \, .
    \label{eq_supMat_lub}
\end{equation}
Derivatives of $v_x$ with respect to the hatted coordinates will be quantities of the same magnitude as the latter are all quantities normalized by their characteristic magnitudes. As such, the second term on the right-hand-side of Eq.\ \eqref{eq_supMat_lub} is by a factor
$(R_0/\overline{\lambda})^2$ smaller than the first term. Provided that $R_0/\overline{\lambda} \ll 1$, we may thus write
\begin{equation}
    (\Delta \mathbf{v} )\cdot \mathbf{e}_x \approx  \frac{1}{r} \partial_r(r \partial_r v_x).
    \label{eq_supMat_lub2}
\end{equation}
In a similar way, one can prove that the variation of pressure in the transversal direction is also by a factor
$(R_0/\overline{\lambda})^2$ smaller than in the longitudinal direction \cite{Antunes2023}, so that one ultimately has $v_r \ll v_x$.
As such, the relevant Stokes equation
(governing fluid flow at small Reynolds numbers) yields
\begin{equation}
    \partial_x P(x,t) =  \mu \frac{1}{r} \partial_r[r \partial_r v_x(x,r,t)], \label{eq:supmat_stokes_lub}
\end{equation}
where $\mu$ is the shear viscosity and where we have made use of the radial symmetry of the problem. 

Equation \eqref{eq:supmat_stokes_lub} can be integrated twice in the radial direction, yielding
\begin{equation}
    v_x(x,r,t) = - \frac{\partial_xP(x,t)}{4 \mu} \left[ R^2(x)-r^2 - \frac{\log\left(\frac{R(x)}{r} \right)}{\log\left(\frac{R(x)}{R_c} \right)} [R^2(x)-R_c^2]\right] + \frac{\log\left(\frac{R(x)}{r} \right)}{\log\left(\frac{R(x)}{R_c} \right)} [ v_p(t) + v_c(x,t)],
    \label{eq_supmat_velprofile}
\end{equation}
when using the boundary conditions in Eqs.\ \eqref{eq_active_bc} and $v_x(x,r=R(x),t) = 0$, as formulated in the main text.

The second term of the right-hand side of Eq.\ \eqref{eq_supmat_velprofile} is proportional to the slip velocity imposed by the ciliary beating and encodes the active driving of the fluid. The first term on the right-hand side is a Poiseuille-like term that is proportional to a local pressure gradient with a local hydrodynamic resistance, which
depends on the thickness of the fluid film. This term is a passive contribution coming from locally-generated pressure gradients that are required to enforce incompressibility, $\nabla \cdot \mathbf{v}(\mathbf{r},t) = 0 $. Indeed, incompressibility implies that the flow rate $Q(t)$ across every cross-section of the film,
\begin{equation}
    Q(t) = 2 \pi \int\limits_{R_c}^{R(x)} r v_x(x,r,t) dr, 
\end{equation}
be homogeneous in space, so it does not depend on $x$.
Plugging in the velocity field from Eq.\ \eqref{eq_supmat_velprofile} yields
\begin{equation}
    Q(t) = f_1(x) \partial_x P(x,t) + f_2(x)[v_p(t) + v_c(x,t)], \label{eq_supmat_Q}
\end{equation}
where
\begin{equation}
    f_1(x) = - \frac{\pi}{8\mu}\left\{ \left( 1 - \frac{1}{\log\left(\frac{R(x)}{R_c}\right)}\right) [R^2(x)-R_c^2]^2 + 2 R_c^2[R^2(x) -R_c^2]\right\}
\end{equation}
and
\begin{equation}
    f_2(x) = 2 \pi \left[ \frac{R^2(x) - R_c^2}{4 \log\left( \frac{R(x)}{R_c}\right)} - \frac{R_c^2}{2} \right]
\end{equation}
are auxiliary functions of position only, which can be computed as both $R_c$ and $R(x)$ are known parameters of the system. Isolating the pressure gradient in Eq. \eqref{eq_supmat_Q} yields
\begin{equation}
    \partial_x P(x,t) = \frac{Q(t) - f_2(x) [v_p(t) + v_c(x,t)]}{f_1(x)}, \label{eq_supMat_pressure}
\end{equation}
which we average now over the entire domain. To do so, we integrate over the domain $x\in [-L,L]$ where $L$ is a generic length and divide the result by $2L$. 
We then take the limit $L\rightarrow \infty$. From here on, we denote this operation on a generic function $g(x)$ with angular brackets such that
\begin{equation}
\langle g(x) \rangle \equiv \lim\limits_{L\rightarrow \infty} \frac{1}{2L} \int\limits_{-L}^L g(x) dx . 
\end{equation}
Averaging Eq. \eqref{eq_supMat_pressure} and using the fact that there is no externally-imposed pressure gradient,
\begin{equation}
    \lim_{L\rightarrow \infty} \frac{P(x=L,r,t) - P(x=-L,r,t)}{2L} = 0, \label{eq_no_superimp}
\end{equation}
yields 
\begin{equation}
    Q(t) = \frac{ \langle \frac{f_2(x)}{f_1(x)}[v_p(t) + v_c(x,t)] \rangle}{\langle f_1^{-1}(x)\rangle}, \label{eq_supmat_Q2}
\end{equation}
where we have used Eq. \eqref{eq_no_superimp}.  The flux $Q(t)$ cannot yet be computed as we do not know $v_p(t)$. We solve this issue 
by using the force-free condition, $ \oint_S \boldsymbol{\sigma}(\mathbf{r},t) \cdot \mathbf{n}(\mathbf{r},t) dS = \mathbf{0}$.
The force per unit length $f_x(L)$ that the fluid exerts on the ciliate in the domain $x\in[-L,L]$ is 
\begin{equation}
    f_x(L) = \frac{2\pi R_c}{2L}  \int\limits_{-L}^L \sigma_{x,r}(x,r=R_c,t) dx \label{eq_supmat_forceLength}
\end{equation}
where $\sigma_{x,r}(x,r=R_c,t)$ is 
\begin{equation}
    \sigma_{x,r}(x,r,t) = \mu [ \partial_x v_r(x,r=R_c,t) + \partial_r v_x(x,r=R_c,t) ].
\end{equation}
The first term in the right-hand side is zero due to the boundary condition in Eq.\ \eqref{eq_active_bc} in the main text. Because the motion is force-free, the total force on the ciliate is zero, and thus so is the force per unit length
\begin{equation}
    \lim\limits_{L \rightarrow \infty} f_x(L) = 0.
\end{equation}
Plugging in the velocity field of Eq. \eqref{eq_supmat_velprofile} in Eq. \eqref{eq_supmat_forceLength} and taking the limit of $L \rightarrow \infty$ yields
\begin{equation}
    \langle \alpha(x) \partial_x P(x,t) \rangle = - \langle \beta(x)[v_p + v_c(x,t)] \rangle, \label{eq_supmat_avg_dxp}
\end{equation}
where
\begin{equation}
    \alpha(x) = \frac{R_c}{2\mu} - \frac{R^2(x)-R_c^2}{4\mu R_c \log\left( \frac{R(x)}{R_c}\right)}
\end{equation}
and
\begin{equation}
    \beta(x) = - \frac{1}{R_c \log\left( \frac{R(x)}{R_c} \right)}
\end{equation}
are auxiliary functions which like $f_1(x)$ and $f_2(x)$ depend only on position and are computable. Plugging the pressure gradient from Eq. \eqref{eq_supMat_pressure} with the flow rate from Eq.\ \eqref{eq_supmat_Q2} into Eq.\ \eqref{eq_supmat_avg_dxp} yields
\begin{align}
    -\langle \beta(x)[v_p(t) + v_c(x,t)] \rangle &= \langle f_2(x) f^{-1}_1(x)[v_p(t) + v_c(x,t)] \rangle \langle f_1^{-1}(x) \rangle^{-1} \langle \alpha(x) f_1^{-1}(x) \rangle \nonumber \\
    & \enspace - \langle \alpha(x) f_2(x) f_1^{-1}(x) [v_p(t) + v_c(x,t)] \rangle. 
\end{align}
Since $v_p(t)$ does not depend on the coordinate $x$, we may isolate it to obtain 
\begin{equation}
    v_p(t) = \dot{x}_p(t) = - \frac{\langle \beta(x) v_c(x,t) \rangle + \langle f_2(x) f^{-1}_1(x) v_c(x,t) \rangle \langle f_1^{-1}(x) \rangle^{-1} \langle \alpha(x) f_1^{-1}(x) \rangle  - \langle \alpha(x) f_2(x) f_1^{-1}(x) v_c(x,t) \rangle}{\langle \beta(x)\rangle + \langle f_2(x) f^{-1}_1(x) \rangle \langle f_1^{-1}(x) \rangle^{-1} \langle \alpha(x) f_1^{-1}(x) \rangle  - \langle \alpha(x) f_2(x) f_1^{-1}(x)  \rangle}, \label{eq_supmat_lubtheory}
\end{equation}
which is finally computable as all variables on the right-hand side are known. 

For further clarity, we write Eq. \eqref{eq_supmat_lubtheory} as
\begin{equation}
    v_p(t) = -\langle \mathcal{G}(x) v_c(x,t) \rangle, \label{eq_SMwithMobility}
\end{equation}
which is a weighted average of $v_c(x,t)$, with a weight 
\begin{equation}
    \mathcal{G}(x) = \frac{ \beta +  f_2 f^{-1}_1 \langle f_1^{-1} \rangle^{-1} \langle \alpha f_1^{-1} \rangle  -  \alpha f_2 f_1^{-1}}{\langle \beta\rangle + \langle f_2 f^{-1}_1 \rangle \langle f_1^{-1} \rangle^{-1} \langle \alpha f_1^{-1} \rangle  - \langle \alpha f_2 f_1^{-1}  \rangle}, \label{eq_SMMobility}
\end{equation}
where we have omitted the function arguments on the right-hand side for improved readability. 

We have now reduced the full 3D hydrodynamics problem to an ordinary differential equation of the form $\dot{x}_p = \mathcal{F}(x_p,t)$, which can easily be solved numerically. Such procedure was done to generate the data in Figs.\ \ref{fig:effect_R0Rc_R1R0} and \ref{fig:resonance}. Because we choose
$R(x)$ to be a $\lambda_w$-periodic function, so is $\mathcal{G}(x)$, and so we may write
\begin{equation}
    \mathcal{G}(x) = \sum\limits_{m=0}^{\infty} \mathcal{G}_m \cos\left( \frac{2\pi}{\lambda_w}mx \right) + \tilde{\mathcal{G}}_m \sin\left( \frac{2\pi}{\lambda_w}mx \right), \label{eq_fourier}
\end{equation}
where $\mathcal{G}_m$ and $\tilde{\mathcal{G}}_m$ are the $m$-th Fourier coefficients
\begin{equation}
    \mathcal{G}_m = \frac{2}{\lambda_w}\int\limits_{0}^{\lambda_w} \mathcal{G}(x) \cos\left( \frac{2\pi}{\lambda_w} m  \right) dx,
\end{equation}
and
\begin{equation}
    \tilde{\mathcal{G}}_m = \frac{2}{\lambda_w}\int\limits_{0}^{\lambda_w}\mathcal{G}(x) \sin\left( \frac{2\pi}{\lambda_w} m  \right) dx.
\end{equation}
Plugging in the sinusoidal form of Eq. \eqref{eq:slip_velocity} and Eq. \eqref{eq_fourier} into Eq. \eqref{eq_SMwithMobility} yields
\begin{equation}
    v_p(t) = -v_0 - v_1 \frac{\mathcal{G}_n}{2}\cos[k_c x_p(t) + \omega t], \label{eq_SMAdler}
\end{equation}
where $\mathcal{G}_n$ is the Fourier coefficient whose index $n$ corresponds to the metachronal wavelength such that 
$\lambda_w/n = \lambda_c$. As described in the main text, this equation is reducible to the Adler equation, for which the time-averaged 
velocity and the period of motion can be computed analytically. Nonetheless, calculating the coefficient $\mathcal{G}_n$ analytically 
presents a formidable challenge in the general case. Analytical progress can be made in the limits of thin fluid films ($R_0/R_c \ll1$) and shallow corrugation ($R_1/R_0\ll1$). In such a limit, one may expand $\mathcal{G}(x)$ to linear order in both of these dimensionless numbers and obtain $\mathcal{G}_1 = 2R_1/R_0$ as the only non-zero Fourier coefficient in this limit, as stated in the main text.

In particular, since the Fourier series of $\mathcal{G}(x)$ does not contain terms corresponding to wavelengths $\lambda_c >  \lambda_w$, 
no motion is attained in this region of parameter space, as stated in the main text and seen in Fig.\ \ref{fig:resonance}(a). In fact, the Fourier series will only contain a term of wavelength $\lambda_c$ if $\lambda_w$ is indeed a multiple of $\lambda_c$, again as discussed in the main text. 

In order to bring out the essential mechanism of confinement-enabled motion of ciliates, we have performed the simplification of assuming sinusoidal functional forms for $R(x)$ and $v_c(x,t)$. Nonetheless, Eqs. \eqref{eq_SMwithMobility} and \eqref{eq_SMMobility} still hold for generic functional forms. More intricate channel geometries (e.g. corrugated channels whose wavelength changes in space) still result in the Adler-like Eq. \eqref{eq_SMAdler}, owing to the sinusoidal functional form of $v_c(x,t)$. Should both $R(x)$ and $v_c(x,t)$ hold non-sinusoidal functional forms, Eq. \eqref{eq_SMAdler} will be augmented with further terms, one for each wavelength that appears in the Fourier series of both functions. The influence of more intricate functional forms is, however, beyond the scope of the current study.  

\section{Flow morphology with changing $R_1/R_0$}

\label{SM_morph}
\begin{figure}[t]
	\centering
   \includegraphics[width=1.0\textwidth]{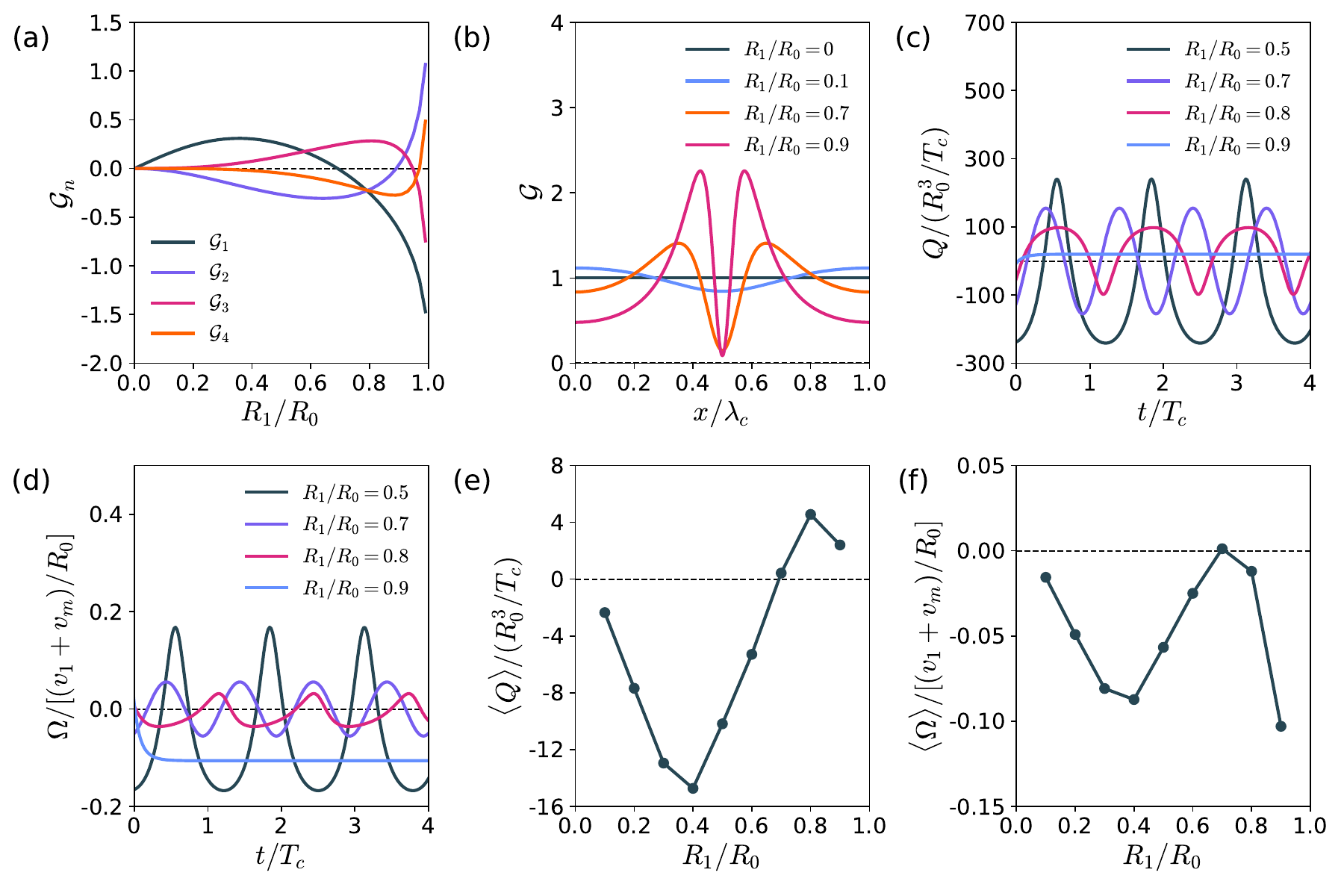}
   \caption{(a) Fourier components $\mathcal{G}_n$ as a function of corrugation amplitude $R_1/R_0$ for various $n$. (b) Mobility function $\mathcal{G}(x)$ as a function of position $x/\lambda_c$ for various $R_1/R_0$. (c) Flow rate $Q$ and (d) vorticity $\Omega$ as a function of time for various $R_1/R_0$. (e) Time-averaged flow rate $\langle Q \rangle $ and (f) vorticity $\langle \Omega \rangle$ as a function of $R_1/R_0$. For all panels, $R_0/R_c=1$, $v_1/v_m=5$,  $v_0=0$, $\lambda_c/\lambda_w=1$.
   }
    	\label{fig:SM_vort}
\end{figure}

The time-averaged velocity (when $v_0=0$ and $\lambda_w = \lambda_c$) was shown in the main text in Fig.\ \ref{fig:effect_R0Rc_R1R0}(b) to have a local minimum upon a change of corrugation amplitude $R_1/R_0$ at a value $R_1^*/R_0 \approx 0.6-0.7$. An analysis of the trajectories reveals that not only the time-averaged velocity is zero, but so is the instantaneous velocity. According to Eq.\ \eqref{eq_withMobility}, such can only occur if $\mathcal{G}_1 =0$. For brevity, we focus on the case $R_0/R_c=1$, $v_1/v_m=5$, and show in Fig.\ \ref{fig:SM_vort}(a) that $\mathcal{G}_1(R_1/R_0)$ indeed displays a root at $R_1^*/R_0 \approx 0.7$. Note that flipping the sign of $\mathcal{G}_1$ results in the same value of $\langle v_p \rangle$, as the 
factor $-1$ may be absorbed in Eq. \eqref{eq_Adler} by shifting the phase $\xi$ by $\pi$. For further clarity, we plot the mobility function $\mathcal{G}(x)$ for varying values of $R_1/R_0$ in Fig.\ \ref{fig:SM_vort}(b).
In order to shed further light on the mechanism underlying the dip in velocity, we examine the morphology of the velocity field, which we 
quantify via two quantities:
the total flow rate $Q(t)$ across the channel,
\begin{equation}
    Q(t) = 2 \pi \int\limits_{R_c}^{R(x)} r v_x(x,r,t) dr,
\end{equation}
which due to incompressibility, is independent of the value of $x$ for which it is calculated; and the total
vorticity averaged over the volume enclosed by the cylinder and one period of the corrugated by wall
\begin{equation}
    \Omega(t) = \frac{2 \pi}{V} \int\limits_0^{\lambda_w}\int\limits_{R_c}^{R(x)} [\partial_x v_r(x,r,t) - \partial_r v_x(x,r,t)] r dr dx \, ,
\end{equation}
where $V$ is the volume
\begin{equation}
   V =  2 \pi \int\limits_0^{\lambda_w}\int\limits_{R_c}^{R(x)} r dr dx \, .
\end{equation}
In Figs.\ \ref{fig:SM_vort}(c) and (d), it can be seen that both quantities either oscillate in time or tend to a steady state, depending on whether the motion is in the oscillatory or phase-locked regime. The time-averaged values of these quantities are non-zero, in general, as can be seen in Figs.\ \ref{fig:SM_vort}(e) and (f). The value of $R_1/R_0$, where both vorticity and flow rate vanish, coincides with $R_1^*/R_0$. Corrugations sufficiently shallower than $R_1^*/R_0$ (e.g. $R_1/R_0=0.5$) showcase large absolute values of $Q(t)$, which result from a layer in which fluid moves along the entire channel. When crossing $R_1^*/R_0$, $Q(t)$ flips sign but attains values much smaller than in the previous case. The vorticity displays the inverse trend, and attains its largest value when $R_1/R_0 > R_1^*/R_0$. The channel-spanning layer becomes then much thinner, and the velocity field is dominated by vortices localized between successive bottlenecks (see also Supplemental Videos 4-6). Motion via the first morphology requires the channel-spanning layer of fluid to travel in the same direction as the ciliate  ($\langle Q \rangle <0$), forcing the vortices near the ciliate to rotate counterclockwise (against the ciliate motion). The second morphology requires the latters to rotate clockwise (with the ciliate motion), forcing the channel-spanning layer to flow in the opposite direction ($\langle Q \rangle >0$, against the ciliate motion). The dip then arises 
when none of these two flow morphologies dominates. Instead, they counteract each other and the ciliate stops moving.

\section{Oblique metachronal wave propagation}

\label{sec:oblique}

\begin{figure}[t]
	\centering
   \includegraphics[width= \columnwidth]{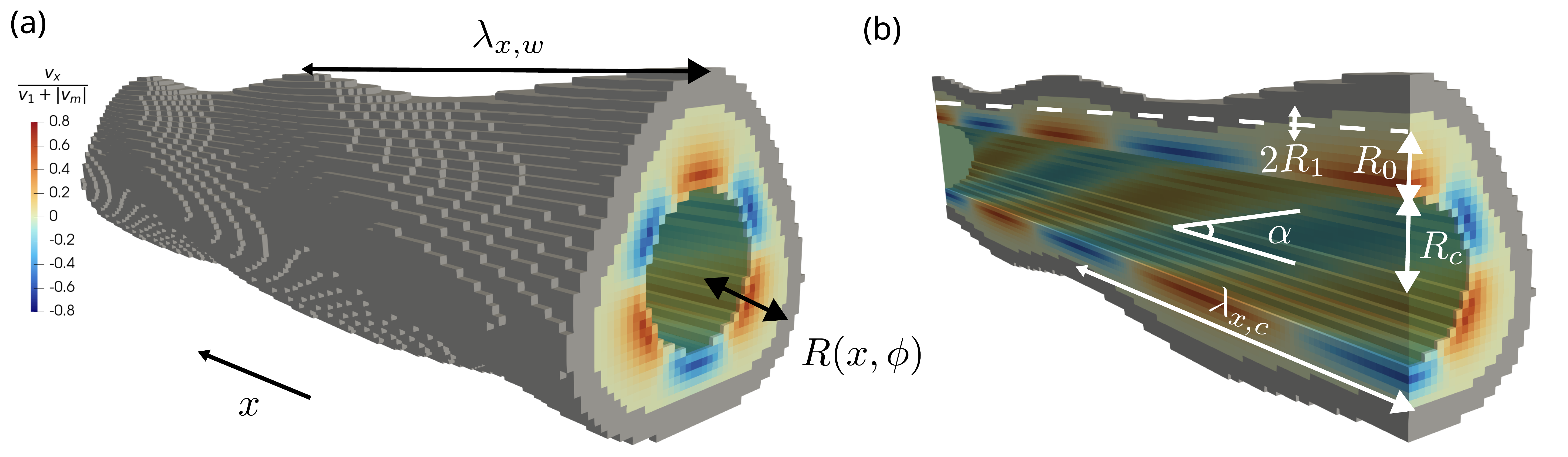}
   \caption{Ciliate with oblique metachronal wave propagation in a rifled channel. The semi-transparent green cylinder represents the ciliate body. The gray shell on the outside represents the rifled channel walls. In-between, there is a layer of fluid that is color-coded to represent the fluid velocity along the $x$ axis. Panel (a) shows the winding grooves running around the exterior and interior channel walls. Panel (b) shows a cross-section where one can see the oblique fronts of the metachronal wave. Parameters: $v_0=0$, $R_0/R_c = 3/5 $, $R_1/R_0=1/3$, $\lambda_{x,c}/\lambda_{x,w} =1$, $R_c/\lambda_{x,c}=1/9$, $v_1/(\omega/k_{x,c})=9.31$, $l_w=l_c$, $\alpha=13.1 \degree$. 
}
    	\label{fig:obliqueModel}
\end{figure}

\begin{figure}[t]
	\centering
   \includegraphics[width= 0.75 \columnwidth]{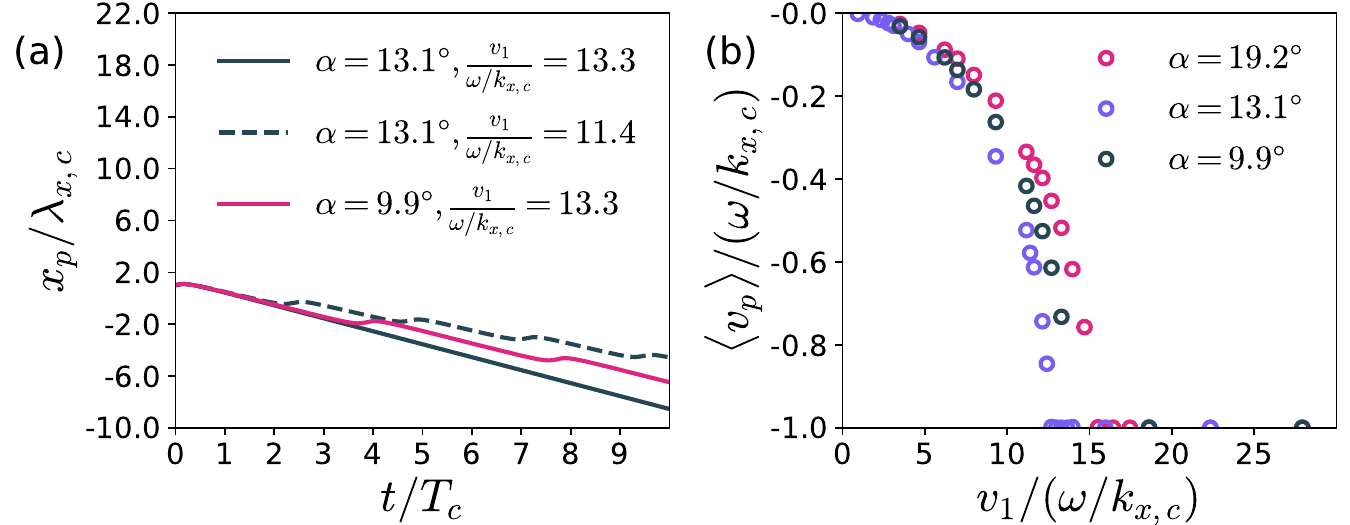}
   \caption{Confinement-induced motion for oblique metachronal wave propagation. (a) Position $x_p$ of the ciliate plotted \emph{vs.} time for different values of $v_1/(\omega/k_{x,c})$ and $\alpha$. Colors indicate value of $\alpha$ and line types indicate $v_1/(\omega/k_{x,c})$.  (b) Time-averaged ciliate velocity $\langle v_p \rangle$ plotted \emph{vs.} $v_1/(\omega/k_{x,c})$,  for several values of $\alpha$. In all panels $\lambda_{x,c} = \lambda_{x,w}$, $l_w=l_c$, $R_0/R_c=3/5$, $R_1/R_0=1/3$, $R_0/\lambda_{x,w} = 1/9$, $v_0=0$. 
}
    	\label{fig:oblique}
\end{figure}

In typical ciliates, the metachronal wave does not propagate along the long axis of the ciliate, but rather in an oblique direction. As such, the metachronal wavefronts are not circles that propagate down the long axis, but helices wrapping around the ciliate which rotate with respect to the long axis (see Fig. \ref{fig:obliqueModel} and Supplemental Video 9). In such a case, the slip velocity condition in Eq. \eqref{eq_active_bc} may be written as
\begin{equation}
    \label{eq:slip_velocity_oblique}
    v_c(x,\phi,t) = v_0 + v_1 \cos\{ k_{x,c} [x -x_p(t)] + k_{\phi,c}R_c\phi - \omega t\},
\end{equation}
where $\phi$ is the azimuthal angle, $x_p(t)$ is the $x$ value of the origin of the frame of reference centered on the body of the ciliate $[x_p(t),0,0]$, and $k_{x,c}$ is the $x$ coordinate of the metachronal wave vector, which now has an azimuthal component such that 
\begin{equation}
    \mathbf{k}_c(\phi) = k_{x,c} \mathbf{e}_x + k_{\phi,c} \mathbf{e}_{\phi}(\phi),
\end{equation}
where $k_{\phi,c}$ is the azimuthal component. From these wavenumbers, we define a longitudinal wavelength $\lambda_{x,c} = 2\pi/k_{x,c}$ and an azimuthal wavelength $\lambda_{\phi,c} = 2\pi/k_{\phi,c}$. To ensure that the phase is continuous, we pick values of $\lambda_{\phi,c}$ which are multiples of the circumference of the ciliate, $\lambda_{\phi,c} = 2\pi R_c / l_c$, where $l_c$ is an integer. Note that the angle $\alpha$ between the wavefronts and the long axis can now be calculated as 
\begin{equation}
    \alpha = \tan^{-1}\left( \frac{\lambda_{\phi,c}}{\lambda_{x,c}}\right),
\end{equation} 
with the case $\alpha=90 \degree$ being discussed in the main text. The constraint on the values of $\lambda_{\phi,c}$ leads to a cross-sectional average of $v_c(x,\phi,t)$ that is always zero. As such, the corrugated channel described in the main text (with $R(x)$ described via Eq. \eqref{eq:Radius}) does not lead to corrugation-induced motion, as there is no difference between what occurs in narrow sections and wide sections of the channel. However, the resonance behavior and the main results described in the main text are recovered when the ciliate is placed in a rifled channel, i.e., with additional corrugations in the azimuthal direction, which spiral in the longitudinal direction, in the same way the metachronal wavefronts do (see Fig. \ref{fig:obliqueModel}). Equation \eqref{eq:Radius} is then generalized as 
\begin{equation}
    \label{eq:RadiusExtend}
    R(x,\phi) = R_c + R_0 + R_1 \cos[k_{x,w} x + k_{\phi,w}(R_c+R_0)\phi],
\end{equation}
where $k_{x,w}$ and $k_{\phi,w}$ are the longitudinal and azimuthal coordinates of the corrugation wavevector
\begin{equation}
    \mathbf{k}_w(\phi) = k_{x,w} \mathbf{e}_x + k_{\phi,w} \mathbf{e}_{\phi}(\phi)
\end{equation}
We introduce the corrugation longitudinal wavelength $\lambda_{x,w} = 2\pi/k_{x,w}$ and an azimuthal wavelength $\lambda_{\phi,w} = 2\pi/k_{\phi,w}$. To ensure that the wall is continuous, we pick values of $\lambda_{\phi,w}$ which are multiples of the circumference of the cylinder of average radius, $\lambda_{\phi,c} = 2\pi (R_c + R_0) / l_w$, where $l_w$ is an integer.

We now pick the parameters of the wall such that $l_w = l_c$, and $\lambda_{x,w} = \lambda_{x,c}$, which are the conditions for resonance between corrugation and the metachronal wave. The 
second equality pertains to resonance in the longitudinal direction, and the 
first pertains to the azimuthal direction. Note that the 
first equality ensures that both the channel radius and the slip velocity profile show the same number of undulations in the cross-section. Lattice-Boltzmann simulations show that the dynamics in the main text ($\alpha=90 \degree$) resurface in this extension to an oblique metachronal wave propagation ($\alpha \neq 90 \degree$): one observes confinement-induced motion when $v_0=0$ due to the resonance between metachronal and corrugation wavevector, as well as the existence of an oscillatory and a steady motion regimes (Fig. \ref{fig:oblique} (a) and Suppementary Movie 9). One can further calculate the time-averaged velocities from the trajectory to see dependencies on, e.g., the parameter $v_1/(\omega/k_{x,c})$, which show functional forms that are qualitatively similar to Fig. \ref{fig:effect_R0Rc_R1R0} (a), as shown in Fig. \ref{fig:oblique} (b). 

We thus demonstrate that the confinement-induced motion described in the main text 
also  exists in the case where the metachronal wave propagates oblique to the main axis, and thus when the system is not radially symmetric.

\section{Supplemental Videos}

There are nine videos supplied with this manuscript:

\begin{itemize}
    \item Supplemental Video 1 depicts motion in the oscillatory regime, as obtained by the theory of section \ref{SM_lub}. 
    Pictured is half of the
    cross-section of the corrugated channel along the longitudinal and radial direction. Black arrows point in the direction of the velocity field and blue/red colors indicate the velocity component $v_x$ in the longitudinal direction. The black area on top corresponds to the channel wall, and the green area to the ciliary body. An eye fixed to the ciliary body has been added. The center of this eye corresponds to the variable $x_p(t)$ and as such shows the motion of the ciliate. Parameters: $v_0=0$, $R_0/R_c = 3/5$, $R_1/R_0=1/3$, $R_0/\lambda_c=3/20$, $\lambda_c/\lambda_w =1$,$v_1/v_m=3.6$.
     
    \item Supplemental Video 2 depicts motion in the phase-locked regime regime, as obtained by the theory of section \ref{SM_lub}. Color coding as per Supplemental Video 1. Parameters: $v_0=0$, $R_0/R_c = 3/5$, $R_1/R_0=1/3$, $R_0/\lambda_c=3/20$, $\lambda_c/\lambda_w =1$, $v_1/v_m=7.5$.

    \item Supplemental Video 3 depicts the motion of a finite ciliate as obtained by lattice-Boltzmann simulations. The ciliate is shaped like a spherocylinder with radius $R_c$ and length $L_c$ (including the hemispherical ends). The slip velocity is only applied on the surface of the cylinder part of the ciliate. In the hemispherical ends, no slip boundary conditions are applied. Color coding as per Supplemental Video 1. Dashed black line marks the eye's starting position. Parameters: $v_0=0$, $R_0/R_c = 3/5$, $R_1/R_0=1/3$, $\lambda_c/\lambda_w =1$, $R_0/\lambda_w=1$, $v_1/v_m=12$, $L_c/\lambda_c = 14.4$, where the length $L_c$ includes the hemispherical ends.

    \item Supplemental Video 4 depicts a ciliate confined to a channel whose value of $R_1/R_0$ is lower than $R_1^\ast/R_0 $, where the dip in Fig.\ \ref{fig:effect_R0Rc_R1R0}(b) occurs. Color coding as per Supplemental Video 1.  Parameters are $R_1/R_0 = 0.5$, $v_1/v_m=5$, $v_0=0$, $R_0/R_c=1$, $R_0/\lambda_w = 1/4$, $\lambda_w/\lambda_c=1$.

    \item Supplemental Video 5 depicts a ciliate confined to a channel whose value of $R_1/R_0$ is approximately that of $R_1^\ast/R_0 $, where the dip in Fig.\ \ref{fig:effect_R0Rc_R1R0}(b) occurs. Color coding as per Supplemental Video 1.  Parameters are $R_1/R_0 = 0.7$, $v_1/v_m=5$, $v_0=0$, $R_0/R_c=1$, $R_0/\lambda_w = 1/4$, $\lambda_w/\lambda_c=1$. 
    
    \item Supplemental Video 6 depicts a ciliate confined to a channel whose value of $R_1/R_0$ is greater than $R_1^\ast/R_0 $, where the dip in Fig.\ \ref{fig:effect_R0Rc_R1R0}(b) occurs. Color coding as per Supplemental Video 1. Parameters are $R_1/R_0 = 0.8$, $v_1/v_m=5$, $v_0=0$, $R_0/R_c=1$, $R_0/\lambda_w = 1/4$, $\lambda_w/\lambda_c=1$. 

      \item Supplemental Video 7 depicts the motion of a finite ciliate as obtained by lattice-Boltzmann simulations. The ciliate is shaped like an ellipsoid with semi-major axis of length $L_m$ and semi-minor axis of length $R_m$. The slip velocity has magnitude described by Eq. \eqref{eq_active_bc}, but points parallel to the ciliate's surface, with zero azimuthal component. Color coding as per Supplemental Video 1. Dashed black line marks the eye's starting position. Parameters: $v_0=0$, $R_0/R_m = 5/6$, $R_1/R_0=1/3$, $\lambda_c/\lambda_w =1$, $R_m/\lambda_w=3/5$, $v_1/v_m=50$, $2L_m/\lambda_c = 16.2$.

     \item Supplemental Video 8 depicts the motion of a finite ciliate as obtained by lattice-Boltzmann simulations. The ciliate is shaped like an ellipsoid with semi-major axis of length $L_m$ and semi-minor axis of length $R_m$. The slip velocity has magnitude described by Eq. \eqref{eq_active_bc}, but points parallel to the ciliate's surface, with zero azimuthal component. Color coding as per Supplemental Video 1. Dashed black line marks the eye's starting position. Parameters: $v_0=0$, $R_0/R_m = 1$, $R_1/R_0=1/3$, $\lambda_c/\lambda_w =1$, $R_m/\lambda_w=3/5$, $v_1/v_m=30$, $2L_m/\lambda_c = 14.4$.

     \item Supplemental Video 9 depicts the motion (oscillatory regime) of a ciliate with a metachronal wave, which propagates oblique to the main axis, as obtained by lattice-Boltzmann simulations. The semi-transparent green cylinder represents the ciliate body. The eyeball is fixed to the ciliate body and shows its motion. The gray channel walls are rifled, showing indentations along the azimuthal direction which twist in the $x$-direction. In between, there is a layer of fluid that is color-coded to represent the fluid velocity along the $x$-axis. Parameters: $v_0=0$, $R_0/R_c = 3/5 $, $R_1/R_0=1/3$, $\lambda_{x,c}/\lambda_{x,w} =1$, $R_c/\lambda_{x,c}=1/9$, $v_1/(\omega/k_{x,c})=9.31$, $l_w=l_c$, $\alpha=13.1\degree$. Definitions for the parameters can be found in Supplemental Material \ref{sec:oblique}. 

\end{itemize}


\end{document}